\documentclass[12pt,preprint]{aastex}

\slugcomment{Submitted to the Astronomical Journal}
\shorttitle{X-ray properties of HE0450-2958} \shortauthors{Zhou, Yang,
L\"u, \& Wang}

\def\xmm{{\it XMM-Newton}}

\def\hst{{\it Hubble}}

\def\h0 {$H_0$=70 km s$^{-1}$ Mpc$^{-1}$}

\newcommand{\he}{HE0450-2958}

\newcommand{\ce}{\ifmmode {\cal E} \else ${\cal E}$\ \fi}
\newcommand{\kms}{\ifmmode {\rm km\ s}^{-1} \else km s$^{-1}$\ \fi}
\newcommand{\ergs}{\ifmmode {\rm erg\ s}^{-1} \else erg s$^{-1}$\ \fi}
\newcommand{\erg}{ergs s$^{-1}$ cm$^{-2}$}
\newcommand{\tes}{\ifmmode \tau_{\rm es} \else $\tau_{\rm es}$\ \fi}
\newcommand{\tk}{\ifmmode \tau_{\rm K} \else $\tau_{\rm K}$\ \fi}
\newcommand{\vfwhm}{\ifmmode V_{\mbox{\tiny FWHM}} \else
            $V_{\mbox{\tiny FWHM}}$\fi}
\newcommand{\msun}{\ifmmode M_{\odot} \else $M_{\odot}$\ \fi}
\newcommand{\afe}{\ifmmode {\mathcal A_{\rm Fe}} \else${\mathcal A_{\rm Fe}}$\ \fi}
\newcommand{\et}{et al.\ }

\newcommand{\ovii}{O {\sc vii}\ }
\newcommand{\oviii}{O {\sc viii}\ }
\newcommand{\lb}{\ifmmode L_{\rm Bol} \else $L_{\rm Bol}$\ \fi}
\newcommand{\ledd}{\ifmmode L_{\rm Edd} \else $L_{\rm Edd}$\ \fi}
\newcommand{\lx}{\ifmmode L_{\rm 2-10keV} \else  $L_{\rm 2-10keV}$\ \fi}
\newcommand{\hb}{H$\beta$}
\newcommand{\mbh}{\ifmmode M_{\rm BH}  \else $M_{\rm BH}$\ \fi}
\newcommand{\lv}{\ifmmode \lambda L_{\lambda}(5100\AA) \else $\lambda L_{\lambda}(5100\AA)$\ \fi}

\received{2006 April }
\begin{document}

\title{X-ray properties of the quasar HE0450-2958}

\author{Xin-Lin Zhou\altaffilmark{1,2}, Fang Yang\altaffilmark{1,2}, Xiao-Rong
L\"u\altaffilmark{1,2}, Jian-Min Wang\altaffilmark{1}}

\altaffiltext{1}{Key Laboratory for Particle Astrophysics, Institute of High Energy Physics,
 Chinese Academy of Sciences, Beijing 100049, China}
\altaffiltext{2}{Graduate School of Chinese Academy of Sciences, Beijing 100049, China}

\begin{abstract}
We present an {\it XMM-Newton} EPIC observation of HE0450-2958 which
may be a ``naked''quasar as suggested by Magain et al.
The \xmm \ EPIC spectra show a substantial soft X-ray
excess, a steep photon index,
 as well as marginal evidence for a weak Fe K$\alpha$ line. 
The X-ray absorption is consistent with the galactic level.
The 0.3-10 keV EPIC spectra can be fitted by a power law plus a
blackbody model, however, the fit by the relativistically blurred photoionized disc
reflection is better. 
We estimate the black hole mass of $2^{+7}_{-1.3} \times 10^{7} M_{\odot}$ from the 
X-ray variability. This broadly agrees with the value derived from the optical \hb~ 
line width. These results support a high-state Seyfert galaxy of the source.
HE0450-2958 shares similar properties of transitionary objects from ultra-luminous 
infrared galaxies to quasars.
We suggest that HE0450-2958 is just in the beginning of an optical
quasar window.

\end{abstract}


\section{INTRODUCTION}
HE0450-2958 is a bright quasar ($z=0.285$) locating $\sim$ $1.5''$ from an 
ultra-luminous infrared galaxy (ULIRG) and revealed by the \hst~ space
 telescope (Boyce et al. 1996; Canalizo \& Stockton 2001).
Recently, there have been controversies on the nature of
HE0450-2958 (Magain et al. 2005; Merritt et al. 2006; Haehnelt et
al. 2006; Hoffman \& Loeb 2006).
HE0450-2958 resides in an elliptical galaxy.  Assuming that the
quasar is accreting at half the Eddington limit with a
radiative efficiency of 10\%, Magain et al. (2005) inferred the mass of
the central supermassive black hole (SMBH) of $8\times10^{8}$
$M_{\odot}$ from the quasar absolute magnitude $M_{V}=-25.8$. They
argued that the host galaxy luminosity is at least 6
times fainter than that predicted
 from the relation between the SMBH mass and the host-galaxy
 spheroid luminosity (Mclure \& Dunlop 2002). They thus concluded that the
 quasar's host galaxy is dark, or the quasar is ``naked''. This suggests the
possible evidence for the ejection of a SMBH from a galaxy merger event
 (Haehnelt et al. 2006; Hoffman \& Loeb 2006).

However, Merritt et al. (2006) presented the optical spectra to argue
 that \he~ is a narrow-line Seyfert 1 galaxy (NLS1). They derived the SMBH mass
 of $(2-11)\times$10$^{7}$  $M_{\odot}$ from the \hb~ line width.
 Then the inferred host-galaxy luminosity from the SMBH mass is consistent
 with the observed luminosity. They also argued that the ejection model 
(Haehnelt et al. 2006; Hoffman \& Loeb 2006) can be ruled out since the narrow emission
line gas remains bound to the SMBH as showed by the quasar's optical spectra.

The controversies lie in the host nature and the SMBH mass of the
quasar. X-ray observations, as a powerful probe of the SMBH activities
(Mushotzky et al. 1993), may provide independent clues to understand
the nature of the source. Here we present an \xmm~ observation of the X-ray
counterpart of \he. We find that it is a high-state Seyfert 1
galaxy, i.e., accreting above the Eddington limit. Throughout this
paper, we use the cosmological parameters of \h0 , $\Omega_{m}=0.3$,
$\Omega_{\Lambda}=0.7$.

\section{DATA REDUCTION}
HE0450-2958 was observed by \xmm \ on 2003 Sep.9 during
orbit 687 (PI: N. Anabuki). The observational details of the
European Photon Imaging Camera (EPIC) onboard \xmm, including the
two MOS cameras (Turner \et 2001) and the pn camera (Str\"uder \et
2001) can be seen in Table 1.

\begin{table}
\footnotesize
\caption{Instrument modes and exposure times. }
\vskip 0.1cm
\begin{tabular}{lcccc}\hline \hline
 Ins. &  Mode & filter & time(ksec) \\
\hline
  MOS 1 & Full Frame & medium & 15.7 \\
  MOS 2 & Full Frame & medium & 15.7 \\
  pn & Full Frame    & medium  & 14.0 \\

\hline
\end{tabular}
\end{table}
\normalsize

The cookbook for the \xmm~ data analysis software SAS in the \xmm~
Data Center at MPE \footnote{{\it
http://wave.xray.mpe.mpg.de/xmm/cookbook}} is referred for the data
reduction. The EPIC data are screened with the SAS v6.0 software
(Gabriel \et 2004), and the corresponding calibration files are
available \footnote{ {\it http://xmm.vilspa.esa.es/ccf}}. The X-ray
events corresponding to patterns 0-4 (single and double pixel
events) for the pn data and patterns 0-12 for the
 MOS data are selected. The EPIC data are used in the 0.3 - 10 keV range and
 hot or bad pixels are removed. We extract the source spectra from a
circle within $38''$(760 pixels) of the detected source position,
with the background being taken from a circular source-free region
with the same size avoiding the CCD chip gaps.
The presence of background flaring in the observation has been checked and 
removed via using a Good Time Interval (GTI) file, leaving 13.1 ks for the pn and
 15.2 ks for the MOS.  We find no pile-ups
in the EPIC data after checking by the SAS task {\it epatplot}. The
response files are generated with the SAS tools {\it rmfgen}  and
{\it arfgen}. Spectral files are binned to at least 20 counts per
bin to apply the $\chi^2$ statistics. Spectral fit is based on the
XSPEC v12.3.0 package (Arnaud 1996). Errors are quoted at the 90\% confidence level ($\Delta
\chi^{2}=2.71$).

The X-ray source is located at $04^{\rm h}52^{\rm m}30^{\rm s}.2$,
$-29^{\circ} 53'34''.6$ (J2000.0). Note that EPIC has a spatial
resolution of $\sim$ $10''$ or worse (Ghizzardi et al. 2001), however, the quasar
 is $\sim$ $1.5''$ apart from
the ULIRG. Thus, the \he~ system can't be resolved spatially by
EPIC. All the fits include the absorption due to the line-of-sight
Galactic column of $N_{\rm H}=1.68\times10^{20}\rm{cm}^{-2}$ (Dickey
\& Lockman 1990), and fitting parameters are given in the
rest-frame.

\section{Results}

\subsection{Temporal analysis}

\subsubsection{excess variance}

We extract the 2-10 keV EPIC pn light curve with the time bin of 256
s, the same as adopted by O'Neill et al. (2005). The count rates show evident 
variations in  $\sim$ ks timescale (Fig. 1).
To quantify the X-ray variability of \he, we invoke the X-ray excess variance 
denoted as $\sigma^2_{\rm rms}$ (Nandra et al. 1997 and Turner et al. 1999),

\begin{equation}
\label{sigma2}
 \sigma^{2}_{\rm rms}=\frac{1}{N\mu^{2}}\sum_{i=1}^{N} [(X_{i}-\mu)^2-\sigma_{i}^{2}],
\end{equation}
where $X_{i}$ is the count rates for the $N$ points in the light curve, with the errors
$\sigma_{i}$. $\mu$ is the arithmetic mean of $X_{i}$.
The errors of  $\sigma^2_{\rm rms}$, which depend on the measurement uncertainties and the
stochastic nature of the source, can be expressed as (O'Neill et al. 2005),
\begin{equation}
\label{sigma2er}
\Delta_{\mathrm{tot}}(\sigma^{2}_{\rm rms})      =      \sqrt{     \left(      \frac{
\sigma_{\mathrm{frac}}  \sigma^{2}_{\rm rms}}{\sqrt{N_{\mathrm{seg}}}}  \right)^{2} +
[\Delta_{\mathrm{boot}}(\sigma^{2}_{\rm rms})]^{2} },
\end{equation}
where $N_{\mathrm{seg}}$ is the number of available ligh-curve segments,
$\sigma_{\mathrm{frac}}$ is a fractional standard deviation, $\sigma_{\mathrm{frac}}=0.74$ for
log$M_{\mathrm{BH}}>$ 6.54 and $\sigma_{\mathrm{frac}}=0.48$ for log$M_{\mathrm{BH}}<$ 6.54.
$\Delta_{\mathrm{boot}}(\sigma^{2}_{\rm rms})$ is the bootstrap uncertainty which comes from
 the bootstrap simulation accounting for the measurement uncertainties.   
We find that $\sigma^2_{\rm rms} = 0.0081\pm0.0063$ from the 2-10
keV  EPIC pn data. We don't use the EPIC MOS data since the MOS data have much
lower count rates with relatively larger errors.

\subsubsection{Time lag}
We calculate the cross-correlation function (CCF) between the light curves of 0.3-2
keV and 2-10 keV with the time bin of 50 s (Fig. 2).
The CCF is obtained using the  {\sc XRONOS} command {\sc crosscor}, which uses a direct
Fourier method to compute the coefficient. Errors are obtained via propagating
the errors of the concerned light curves through the cross correlation formulae.
The results show the hard X-ray to be no lagging with respect to the soft X-ray emission. 

\subsection{Spectral analysis}

\subsubsection{Power law}
We use all the EPIC data (pn+MOS1+MOS2) for the spectral analysis to improve the
  photon statistics allowing tighter constraints on spectral parameters.
Initially we use a power law to fit the data above 1 keV with a free
intrinsic absorption. A good fit can be obtained
($\chi^2_{\upsilon}$ of 1.06 for 521 d.o.f.), with the intrinsic absorption
 consistent with zero. A substantial soft X-ray excess below 1 keV can be seen in
the EPIC pn and MOS data when extrapolating this power law over the
full energy range of EPIC (Fig. 3). 

\begin{table*}[t]
\tiny
\caption{Spectral fits to the combined EPIC (pn+MOS1+MOS2) data}
\begin{tabular}{llcccccccccccr}\hline \hline
Model$^a$   & $\Gamma$            & $N_{\rm H}^{\rm int}$ &
 $kT_{BB}$ & $N_{BB}$ & $E_{edge}$  & $\tau$ & $E_{K\alpha}$ & $EW_{K\alpha}$
  & $\theta$  &log$\xi$   & $\chi^2/d.o.f.$  \\
              &                    &  (10$^{20}$ cm$^{-2}$) &  (eV) &  ($\times$10$^{-5}$) &
              (eV)
&   & (keV) & (eV) & \\
(1)  & (2) & (3) & (4) & (5) & (6) & (7) & (8)& (9) & (10) & (11) & (12) \\

\hline

1.   & $2.13^{+0.04}_{-0.03}$ & $<0.03$ & $99^{-3}_{+2}$ &
$9.7^{+3.4}_{-0.5}$ & -& - &- &- &
-& - & 872.8/747 \\
2.   & $2.16^{+0.03}_{-0.03}$ & $<0.01$ & $107^{+5}_{-8}$ &
$7.5^{+1.7}_{-0.4}$
& $773^{+34}_{-18}$ & $0.29^{+0.07}_{-0.11}$ & - & - &-&- & 845.0/745 \\
3.  & $2.16^{+0.03}_{-0.03}$ &$<0.01$ & $106^{+6}_{-6}$
&$7.7^{+1.4}_{-0.6}$ &
$776^{+29}_{-19}$    &  $0.26^{+0.09}_{-0.07}$ & $6.40^f$ & $<64$ &-&-& 842.2/744 \\
4.  &  $2.15^{+0.04}_{-0.04}$ & $<0.01$ & - & - & $788^{+31}_{-29}$ & $0.38^{+0.08}_{-0.07}$ 
& $6.40^f$ & $<36$ & $42^{+8}_{-10}$ & $3.2^{+0.3}_{-0.1}$ & 798.1/738 \\

5.  &  $2.35^{+0.02}_{-0.02}$ & $<0.01$ & - & - & $759^{+11}_{-13}$
& $0.49^{+0.06}_{-0.07}$
& $6.40^f$ & $<42$ &$38^{+10}_{-8}$ &$3.4^{+0.2}_{-0.2}$ & 849.2/741\\
& &   &  & & $1190^{+30}_{-40}$ & $0.30^{+0.07}_{-0.08}$ &   &  & \\

\hline
\end{tabular}
{\baselineskip 9pt \indent \\
Note. -(1): model; (2): photon index; (3): intrinsic column density;
(4): blackbody temperature; (5)  blackbody normalization; (6) energy
of  absorption \\ edge; (7) optical depth of absorption edge; (8)
energy of Fe K$\alpha$ line; (9) equivalent
 width of Fe K$\alpha$ line; (10) inclination angle in the reflection model \\ of {\sc pexriv}; ]
(11) ionization parameter in {\sc pexriv}.
 $^a$ Model expressed in XSPEC command. 1: phabs * zphabs (powerlaw+zbbody);
2: phabs * zphabs\\ * zedge (powerlaw+zbbody);
 3: phabs * zphabs * zedge (powerlaw+zbbody+zgauss);
4:  phabs * zphabs * zedge (kdblur (powerlaw+atable $\lbrace$reflion.mod$\rbrace$)+zgauss);
5: phabs * zphabs * zedge * zedge (pexriv+zgauss). $^f$ fixed.}
\end{table*}
\normalsize

We use the models listed in Table 2 to fit 0.3 - 10 keV spectrum.
 The soft X-ray excess is traditionally taken as thermal emission (e.g. Pounds et al. 1995).
Model 1 is an absorbed power law plus a blackbody model. This gives
an acceptable fit over the full energy range ($\chi^2_{\upsilon}$ of 1.17 
for 747 d.o.f., Model 1 in Table 2), with the photon index of $2.13^{+0.04}_{-0.03}$. 
There are some evidences for an absorption gap between
0.7-0.8 keV in the spectra.  We add an absorption edge to Model 1.
This improves the fit significantly ($\Delta \chi^2$ of 27.8 for 2 fewer d.o.f., Model 2). 
We find that the edge significance is at the level of $>99.9\%$ by applying the
$F$-test. It should be due to the absorption edge of \ovii in 739 eV
(Reynolds 1997). The second absorption edge due to \oviii is not
required in this model ($\Delta \chi^2$ of 1.2 for 2 fewer d.o.f. when including the
 second absorption edge).

\subsubsection{Fe K line}
We first try to add a Gaussian line with all free parameters. However, 
this fit does not improve significantly ($\Delta \chi^2$ of 2.7  for 3 fewer d.o.f.).
Since the Fe K$\alpha$ lines in current data are generally narrow (Nandra
2006), we fix the intrinsic width of the line at 10 eV. This fit is still 
poor ($\Delta \chi^2$ of 2.6  for 2 fewer d.o.f.), and the line energy can not be constrained
well. We also fix the line energy at 6.40 keV. This returns an equivalent width of $<64$ eV. 
  The line significance $F^{line}$ is at a level of 89\% compared to the continuum alone.
Generally, the significance should be larger than 90\% for the presence of
 a Fe K$\alpha$ line. Thus, the line is barely significant.

\subsubsection{soft X-ray excess}

Recently, it has been suggested that the soft X-ray excess can be
arised from a relativistically blurred photoionized disc reflection
in a large sample of AGNs (Crummy et al. 2006). The relativistic
convolution {\sc kdblur}  has been included in XSPEC v12.3.0 and 
the photoionized disc reflection model of {\sc REFLION} (Ross \& Fabian 2005) is also available
\footnote{{\it http://heasarc.gsfc.nasa.gov/docs/xanadu/xspec/models/reflion.html}}.
We follow Crummy et al. (2006), fix the outer radius of accretion disc at 100 $R_{\rm g}$.
This fit (Model 4) is better than the blackbody fit of Model 3 ($\Delta \chi^2$ of 44.1 
for 6 fewer d.o.f., see Table 2), with the
 inner radius of the accretion disc at $1.6^{+0.6}_{-0.2}$ $R_{\rm g}$, the index of the
 emissivity of the accretion disc of $8.1^{+1.9}_{-3.5}$, the iron abundance of 
the accretion disc of $0.4^{+0.1}_{-0.2}$ (relative to solar), the photon index for 
illuminating power-law spectrum of $2.38^{+0.03}_{-0.02}$. Other parameters in this model have been listed in Table 2.

We further test the reflection scenario by using the {\sc pexriv}
model (Magdziarz \& Zdziarski 1995) to fit 0.3-10 keV spectrum.
Although the {\sc pexriv} model doesn't include the relativistically blurring effect, it can
 generally represent the
reflection from the ionized material. We fix the high-energy cutoff at 200 keV, the reflector
 at unity (Malizia et al. 2003). This fit is worse than the blackbody model 
($\Delta \chi^2$ of -7 for 3 fewer d.o.f. ), but still acceptable ($\chi^2_{\upsilon}$ of 1.15
 for 741 d.o.f.).

\section{Discussion}

\subsection{BH mass}
 We calculate the black hole (BH) mass based on the
anti-correlation between the X-ray excess variance $\sigma^2_{\rm rms}$ and the BH
mass (Lu \& Yu 2001; Bian \& Zhao 2003; Papadakis 2004; O'Neill et al. 2005).
Using the correlation derived from Eq. 3 in O'Neill et al. (2005),
\begin{equation}
\log M_{\rm BH}=5.75+1.20\log \left( \frac{0.144}{\sigma^2_{\rm rms}}-1\right),
\end{equation}
we obtain the BH mass of $2^{+7}_{-1.3}\times10^{7}$ $M_{\odot}$ (Fig. 7).

We plot the spectral energy distribution (SED)
 for \he~ in Fig. 8. There is a ``Big Blue Bump''(BBB) in
 the UV band. We fit the BBB with a standard thin accretion
disk model (D\"oerrer et al. 1996) and fit the X-ray data with a power law plus
a blackbody model. The accretion disk fit (dotted line in Fig. 8) returns a
large BH mass of $8\times10^{8}$ $M_{\odot}$, with the
 Eddington ratio of 0.3, the inclination angle cos$\theta=0.5$ and the BH spin parameter
 $a=0.6$. This BH mass is much
higher than our result. It is possible that the UV and optical data is
 contaminated by the nearby ULIRG and thus the SED luminosity is overestimated.
A similar case appears in NLS1 Ton S180, 
 the estimation of BH mass based on the SED from the
 simultaneous multiple band observations is also much higher than that from the
 \hb~ line width (Turner et al. 2002). Alternatively, it is not a reliable way to
estimate the BH masses in NLS1s by fitting SED using a standard  thin disk model,
since they may accrete above the Eddington limit and the standard thin disk model
does not work in this case (Kawaguchi et al. 2004). It is worth studying the SED of
\he~ through more sophisticated slim disk model (Abramowicz et al.
1988; Wang et al. 1999) in future.

\subsection{Eddington ratio}

 The X-ray excess variance is related with the observed X-ray luminosity
$L_{\rm X}$ and the Eddington ratio ${\cal E}\equiv L_{\rm bol}/L_{\rm Edd}$
 through (Leighly 1999),
\begin{equation}
\sigma^2_{\rm rms}\propto \left( \frac{L_{\rm X}}{\eta {\cal
E}}\right)^{1-\alpha},
\end{equation}
where  $L_{\rm bol}$ is the bolometric luminosity and $L_{\rm Edd}$
is the Eddington luminosity, $\eta$ is the radiation efficiency and
$\alpha$ is the slope of the power spectrum, assuming $\alpha=2$
(Leighly 1999). The integrated $2-10$ keV flux of \he~ is
$2.00\times 10^{-12}$ \erg, corresponding to a luminosity of 5.13
$\times$ 10$^{44}$ ergs s$^{-1}$. We find that \he~ is located in
the NLS1 region in the $\sigma^2_{\rm rms}-L_{\rm X}$ plot of
Leighly (1999), indicating that \he~ has a high Eddington ratio.
This supports the hypothesis that HE0450-2958 is a NLS1 galaxy.

The soft and the hard X-rays in \he~ can arise from very close
regions. As suggested above, a slim disk may be powering \he.
 The fluctuations of the slim accretion disk can be the
origin of the simultaneous X-rays variations (Mineshige et al. 2000;
Wang \& Netzer 2003). In this case, the slim disk suffers the
so-called ``photon bubble instability'' (Gammie 1998) since the
energy densities of the trapped photons and the magnetic field are
larger than those in standard accretion disks. This leads to the
hard X-rays closely following the soft X-rays variations. The
zero lag between the soft and the hard X-rays in \he~ supports this scenario.

The yielded hard photon index of $2.16^{+0.03}_{-0.03}$, is
steeper than the average photon index of $1.8-2.0$ in Seyfert 1
galaxies (Nandra \& Pounds 1994; George et al. 2000) and in quasars
(Reeves \& Turner 2000). The AGNs with steep X-ray indices as
 well as the strong soft X-ray emission have been considered as high-state objects
 (Pounds et al. 1995).
According to the anti-correlation between the FWHM (\hb) and the
hard X-ray index (Brandt et al. 1997), the FWHM (\hb) of \he~
should be narrow, also consistent with a NLS1.

The hard X-ray photon index is related to Eddington ratio 
(e.g. Shemmer et al. 2006). We calculate ${\cal E}$ from 
the $\Gamma_{\rm 2-10 keV}- {\cal E}$ relation in Wang et al. (2004),
\begin{equation}
\Gamma_{\rm 2-10 keV}=2.05+0.26\log{\cal E}.
\end{equation}
We find ${\cal E}= 3^{+1}_{-1}$ for $\Gamma_{\rm 2-10
keV}=2.16\pm0.03$. This suggests that \he~ is accreting above the Eddington
limit.

The AGN bolometric luminosity can be estimated from the X-ray
luminosity by multiplying the
 bolometric correction $f_{\rm bol/x}$. We
find the bolometric luminosity to be $9^{+16}_{-6}\times$ 10$^{45}$
ergs s$^{-1}$ by assuming $f_{\rm bol/x} = 17^{+33}_{-11}$.  The $f_{\rm bol/x}$ we used is 
the range in conversion factors that the bolometric luminosity accounts
for the X-ray luminosity found in Elvis et al. (1994), also
 consistent with recent estimation by Marconi et al. (2004) and Barger et al. (2005).
 We then estimate the BH mass of $2^{+5}_{-1.2}$$\times$
10$^{7}$ $M_{\odot}$ for $L_{\rm bol}=9^{+16}_{-6}\times$ 10$^{45}$ ergs s$^{-1}$ and 
 ${\cal E}= 3$, agreeing with the $M_{\rm BH}$ from the X-ray variability.

 The weak
Fe K$\alpha$ emission can be also due to the high Eddington ratio
(Pounds et al. 2003) for the anti-correlation between the equivalent width of
narrow Fe K$\alpha$ line and Eddington ratio (Zhou \& Wang 2005).

\subsection{On the origin of Soft X-ray excess}
The origin of the soft X-ray excess is still under debate.
The traditional view is that the soft X-ray excess is the high-energy tail of BBB
 which is the thermal emission from the inner region of the accretion disc.
The inferred Eddington ratio (${\cal E}= 3^{+1}_{-1}$), is too high
for a standard accretion disk. The slim disc can be
applied for the super-Eddington accretion.
 The effective temperature of the slim disc can be written as (Wang \& Netzer 2003),
\begin{equation}
T_{\rm eff}= 3.14 \times 10^{3} \gamma_{0}^{-1/4}M_{\rm BH}^{-1/4}\left(r/r_{\rm s}\right)^{-1/2} {\rm eV},
\end{equation}
where $\gamma_{0}=(5+\alpha^{2}/2)^{1/2}$ is a weak function of the
viscosity $\alpha$, $\gamma_{0}\approx2.24$ for  $\alpha\ll1$. We
find $T_{\rm eff}\approx $ 23 eV for $r=3 r_{\rm s}$ and $M_{\rm
BH}= 2\times10^7 M_{\odot}$. However, the derived blackbody
temperature from the spectral fitting, $kT_{BB}=106\pm6$ eV,
is much higher than expected. For the thermal origin of
the soft X-ray excess, it is difficult to reconcile the BH mass to the 
blackbody temperature (e.g., a slim disc temperature 100 eV requires a BH mass
 of $<10^5 M_{\odot}$). However the innermost region of accretion disk may be very
 complicated, such as strong outflow developed from the disk itself, evidenced by
 PG 1211+143 (Pounds \& Page 2006). Comptonization inside the outflow has not been studied,
 but might significantly contribute to the soft X-ray excess. Actually the temperature holds
a constant in the transition layer between the hot corona and cold disk, it may also have
 important contribution to the soft X-ray excess (Nayakshin \& Melia 1997). As noted by
 Gierli\'nski \& Done (2004), many AGNs with quite 
different BH masses have soft X-ray excess with the temperature confined in a very narrow range.
This points out towards an atomic nature (either related to absorption or reflection)
for the soft excess, rather than a thermal origin.

 Whereas the blackbody model can fit the spectra,
 our result shows that the fit by relativistically blurred
 disc reflection model is better. This supports that
 the soft X-ray excess can arise from the disc reflection.
The ionization parameter given by  the reflection model 
is large (log$\xi=3.2^{+0.3}_{-0.1}$), implying that the reflection material
 is highly ionized.
The simulation of X-ray photoionized accretion disc shows that the surface of the
accretion disc can be significantly ionized at a high Eddington
ratio (Matt et al. 1993). The highly-ionized disc surface becomes reflective
 in the soft X-ray band,
 producing a steepening X-ray continuum (Haardt \& Maraschi 1993; Nayakshin et al. 2000;
Ballantyne et al. 2001). This scenario is physically plausible for \he.

\subsection{X-ray properties of the transitionary objects}

Canalizo \& Stockton (2001) compiled a sample (including \he) of
low-redshift ($z\leqslant0.4$) objects that are
 in a transitionary stage between ULIRGs and quasars. Their sample selected from
 the intermediate position in the far-infrared color-color diagram between the regions
occupied by the two classes of objects from the IRAS all-sky survey is nearly complete.
We collect the X-ray data of Canalizo \& Stockton (2001) sample given in Table 3.
 For seven X-ray detected objects, all show steep X-ray photon
indices and small X-ray intrinsic absorption, with the exception of
 Mrk 231, which is heavily obscured. Thus \he~ is very similar to other objects.

\clearpage

\begin{table}
\footnotesize
\caption{The transitionary objects in Canalizo \& Stockton (2001) }
\begin{tabular}{llcccc}\hline \hline
 Name &  $z$       & log$L_{\rm IR}$ &  $\Gamma_{\rm 2-10 keV}$ &
$N_{\rm H}^{\rm int}$ & Ref. \\
      &          & ($L_{\odot}$)  &          & (10$^{22}$  cm$^{-2}$)      \\
\hline
I Zw1      & 0.061 &  11.97   &  $2.31^{+0.03}_{-0.03}$ & $0.09^{+0.02}_{-0.02}$ & 1 \\
3C48       & 0.367 & 13.02    &  $1.96^{+0.04}_{-0.04}$  & $<0.21$ & 4          \\
IR 07598+6508 &  0.148 & 12.54 & $2.9^{+0.6}_{-0.5}$     & $0.08^{+0.1}_{-0.06}$ & 5  \\
Mrk 231    & 0.042 &  12.55    & $2.48^{+0.20}_{-0.11}$  & $265^{+173}_{-85}$  & 6 \\             F00275-2859 & 0.279&  12.71    &   -                      &  -         &  -           \\
PG 1700+518 & 0.292 & 12.70    &   X-ray                & undetected              &       3        \\
HE 0450-2958 & 0.285           &12.72 & $2.16^{+0.03}_{-0.03}$  & $<0.01$ & 2  \\
PG 1543+489   & 0.400 & 12.78  &  $2.64^{+0.19}_{-0.20}$ & $0.07^{+0.11}_{-0.07}$  & 3  \\

Mrk 1014   & 0.163 & 12.63     &  $2.24^{+0.06}_{-0.07}$ & 0. & 1    \\

\hline
\end{tabular}
\vskip 0.1cm
\parbox{3.05in}
{\baselineskip 9pt \indent { Ref. for X-ray data: 1. Piconcelli et
al. (2005); 2. this work; 3. George et al. (2000); 4. Siemiginowska
et al. (2003); 5. Imanishi \& Terashima (2004); 6. Braito et al.
(2004). } }

\end{table}
\normalsize

In the hierarchical formation paradigm,  quasars are formed and
fueled via galaxy-galaxy mergers (Kauffmann \& Haehnelt 2000; Di
Matteo et al. 2005). During the major merger of two comparable
galaxies, the quasar is dust-enshrouded and the SMBH growth is
obscured by the gas funneled toward a merger nucleus.
 This picture is supported by a new population of submm and hard X-ray sources at
 $z\approx1.5-3$ (Chapman et al. 2003; Alexander et al. 2005).
 When SMBH reaches a critical mass, the feedback from the SMBH
 activity expels gas, and cleans the obscuring material (Silk \& Rees 1998;
Fabian 1999; Ciotti \& Ostriker 2001). The detailed simulation of
galaxy mergers shows that this process creates a window in which
 the SMBH is observable as an optical quasar for the duration of  $\sim10-20$ Myr for
 a $B$-band luminosity greater than 10$^{11}$ $L_{\odot}$ (Hopkins et al. 2005). We
 argue that \he~ is just in the beginning
of the optical quasar window for: 1) the small X-ray
intrinsic absorption, implying that the AGN is dust-cleaned and
optical-visible; 2) the high accretion rate inferred from the steep
X-ray index and the NLS1 nature; 3) the relatively smaller BH mass
compared with the optical-selected quasars (Hao et al. 2005;
Kawakatu et al. 2006).

The IR luminosities of ULIRGs denote the intense bursts of star
formation and also the ``violence'' of interactions and mergers
(Sanders \& Mirabel 1996). The steep X-ray photon indices of AGNs
denote the super-Eddington accretion of SMBHs. These ULIRGs
associated with AGNs with steep X-ray indices shed new light on the
coeval growth of the stellar bulges and SMBHs in the hierarchical
paradigm.

\section{Conclusions}
The \xmm \ EPIC spectra of \he~ show a substantial soft X-ray
excess, a steep photon index,
 as well as marginal evidence for a weak Fe K$\alpha$ line. 
The X-ray absorption is consistent with the galactic level.
 The 0.3-10 keV EPIC spectra 
can be fitted by a power law plus a
blackbody model, however, the fit by the relativistically blurred photoionized disc
reflection is better. 
We estimate the black hole mass of $2^{+7}_{-1.3} \times 10^{7} M_{\odot}$ from the
 X-ray variability.
This broadly agrees with the value derived from the optical \hb~ line width. 
These results support a high-state Seyfert galaxy of the source.

\he~ shares similar properties of transitionary objects from ultra-luminous infrared
galaxies to quasars. we suggest that \he~ is just in the beginning of the optical quasar
 window.

\acknowledgements{We are grateful to an anonymous referee for the constructive comments
and help with English. W.-H. Bian, Z.-H. Fan,  F. Zhang and Y.-M. Chen are thanked 
for reading the manuscript. We also appreciate the discussion among people
in IHEP AGN group.  This research is supported by NSFC through NSFC-10325313,
10233030 and 10521001.}

\clearpage

\begin{figure}
\figurenum{1}
\centerline{\includegraphics[angle=-90,width=\textwidth]{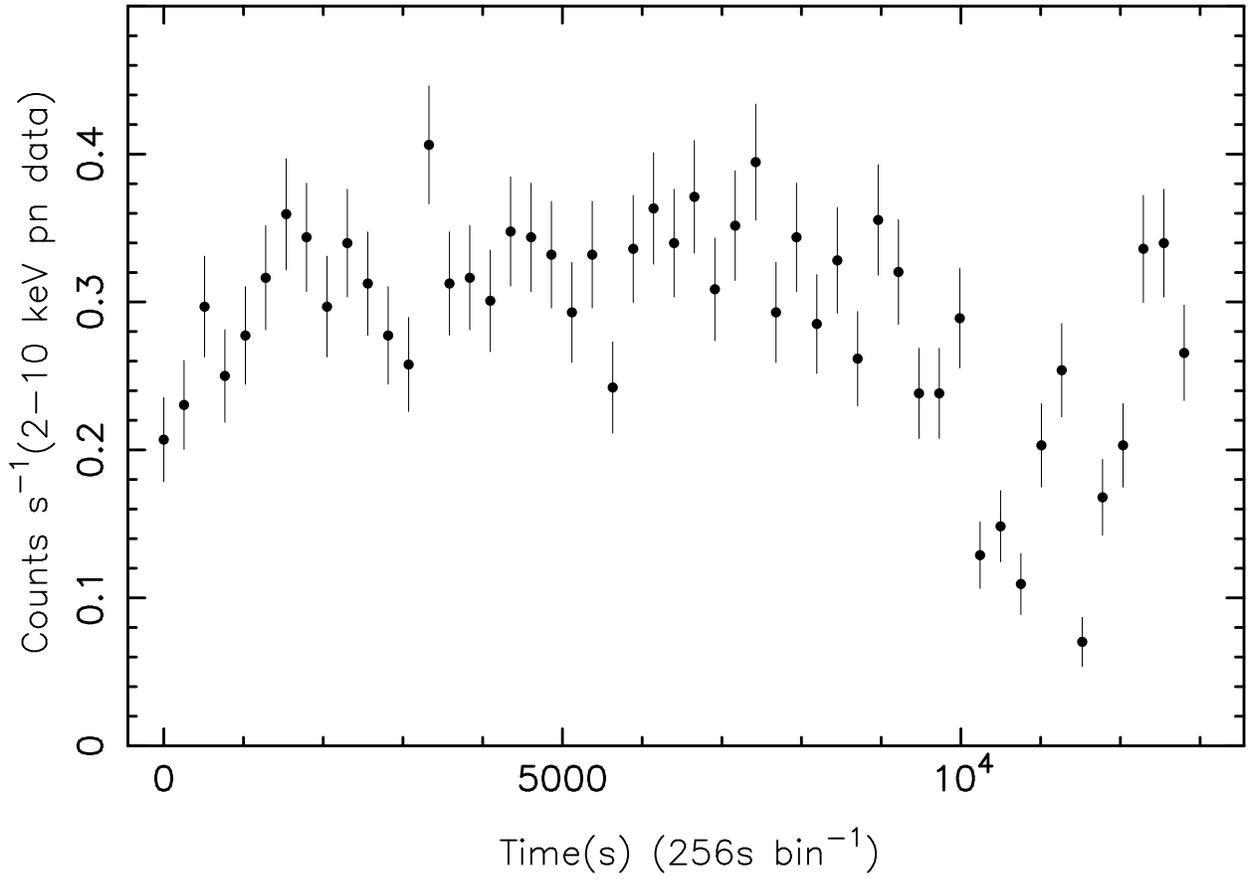}}
\figcaption{The 2-10 keV EPIC pn light curve of \he~ with the time bin
of 256 s shows the rapid variability. }
\label{fig1}
\end{figure}
%

\clearpage

\begin{figure}
\figurenum{2}
\centerline{\includegraphics[angle=-90,width=\textwidth]{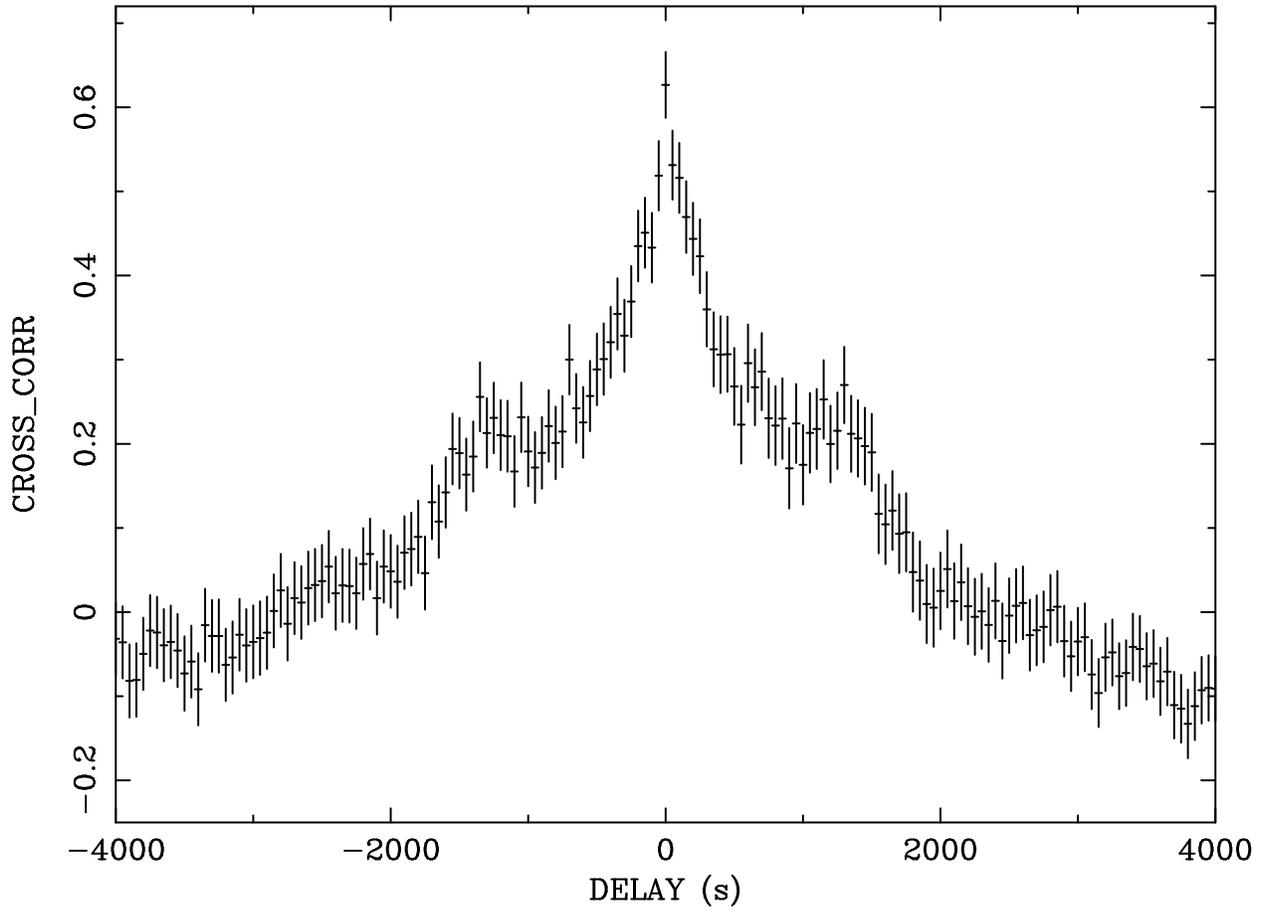}}
\figcaption{The cross-correlation function between the light curves of 0.3-2
keV and 2-10 keV with the time bin of 50 s shows no time lag
 between these two X-ray bands. }
\label{fig2}
\end{figure}
%

\clearpage

\begin{figure}
\figurenum{3}
\centerline{\includegraphics[angle=-90,width=\textwidth]{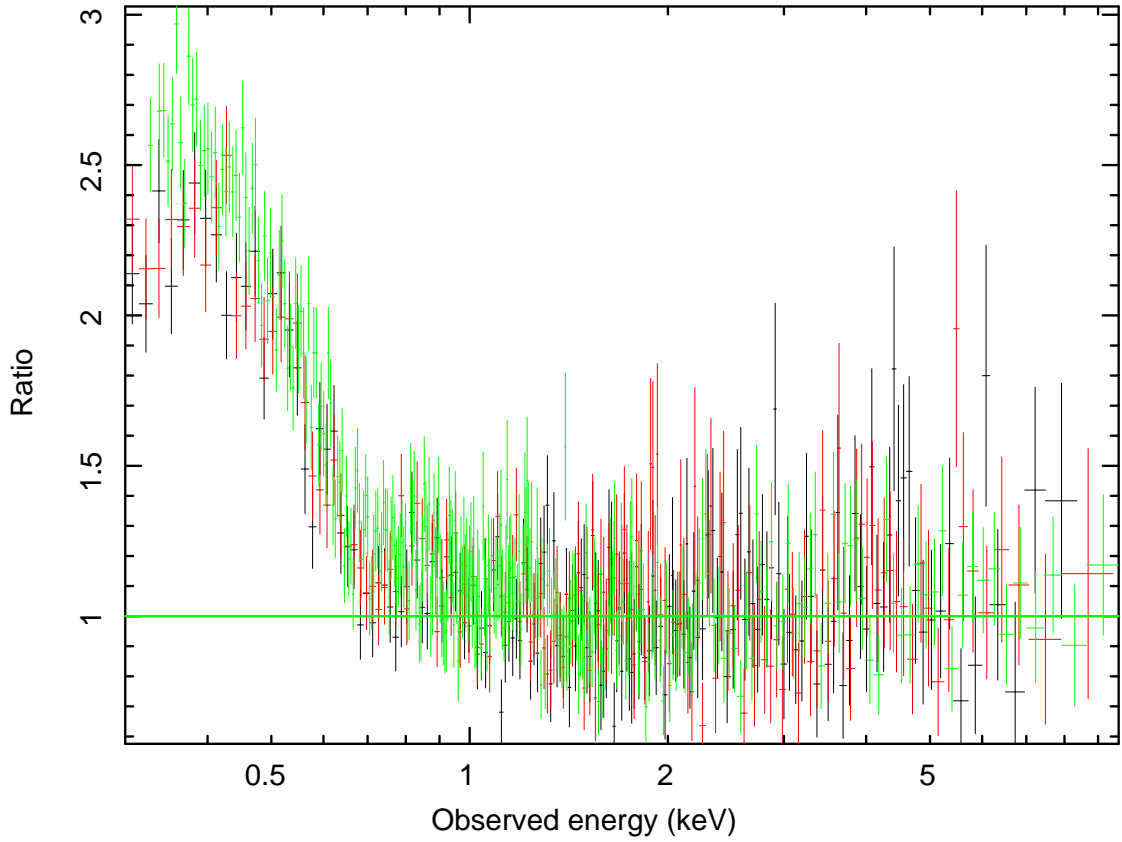}}
\figcaption{The EPIC pn (green), MOS1 (black) and MOS2 (red) spectra of
\he. The data is fitted with a simple power law over 1-10 keV, index
$\Gamma=2.14\pm0.03$. A soft X-ray excess is clear seen below 1 keV.}
\label{fig3}
\end{figure}
%

\clearpage

\begin{figure}
\figurenum{4}
\centerline{\includegraphics[angle=-90,width=\textwidth]{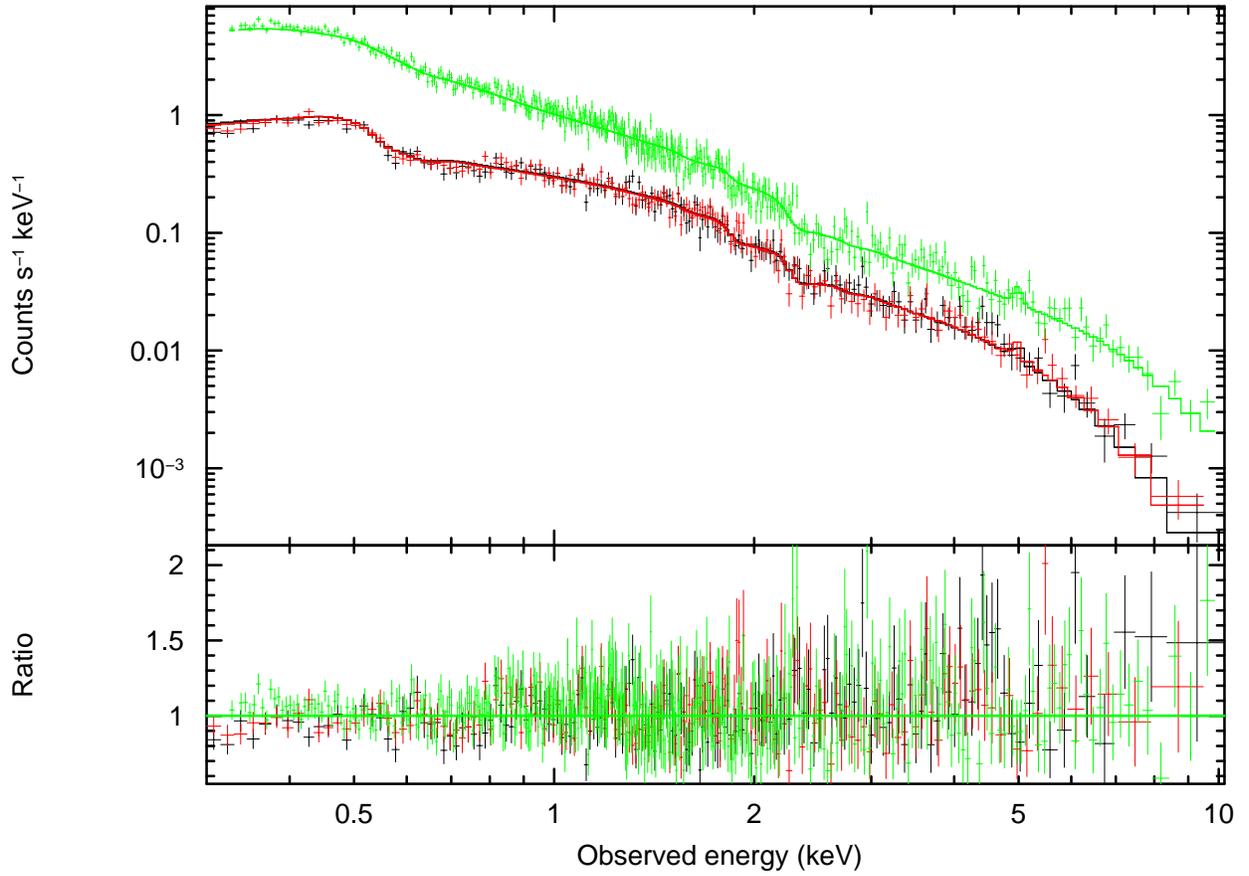}}
\figcaption{The power law plus blackbody model fit to the EPIC pn (green), 
MOS1 (black) and MOS2 (red) data ($\chi^2_{\upsilon}$ of 1.13 for 744  d.o.f., 
Model 3 in Table 2). This model also includes a narrow Gaussian and an absorption edge. }
\label{fig4}
\end{figure}
%

\clearpage

\begin{figure}
\figurenum{5}
\centerline{\includegraphics[angle=-90,width=\textwidth]{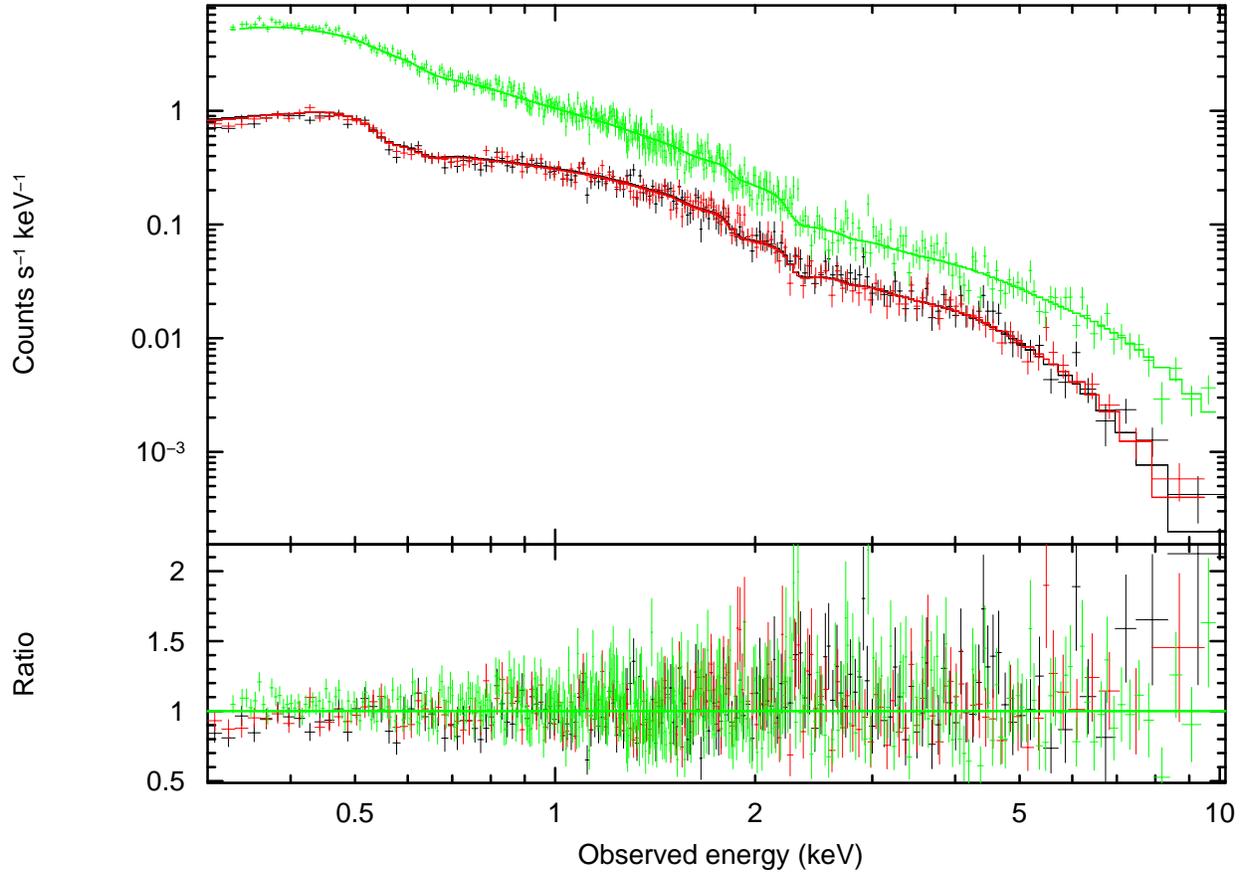}}
\figcaption{The relativistically blurred photoionized disc reflection fit 
to the EPIC pn (green),  MOS1 (black) and MOS2 (red) data (Model 4 in Table 2). 
This model also includes a narrow Gaussian and an absorption edge.
The fit is statistically better than the blackbody fit 
($\Delta \chi^2$ of 44.1 for 6 fewer d.o.f.) and physically plausible. }
\label{fig6}
\end{figure}
%
\clearpage

\begin{figure}
\figurenum{6}
\centerline{\includegraphics[angle=-90,width=\textwidth]{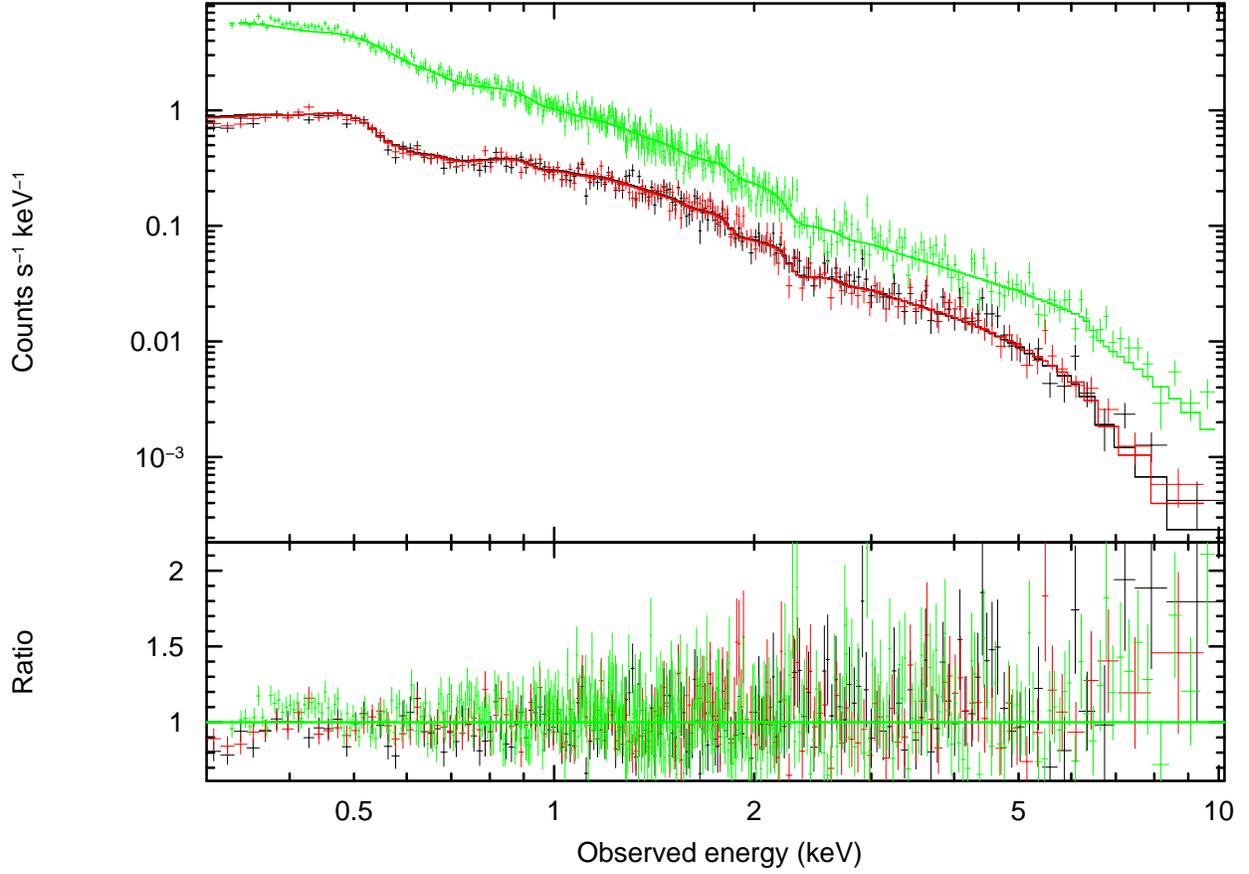}}
\figcaption{The reflection model of {\sc  pexriv} fit to the EPIC pn (green), 
MOS1 (black) and MOS2 (red) data (Model 5 in Table 2). This model also includes
two absorption edges and a  narrow Gaussian. The fit is worse than
the blackbody fit ($\Delta \chi^2$ of -7 for 3 fewer d.o.f. ), 
but still acceptable ($\chi^2_{\upsilon}$ of 1.15 for 741 d.o.f.).}
 \label{fig6}
\end{figure}
%

\clearpage

\begin{figure}
\figurenum{7}
\centerline{\includegraphics[angle=-90,width=\textwidth]{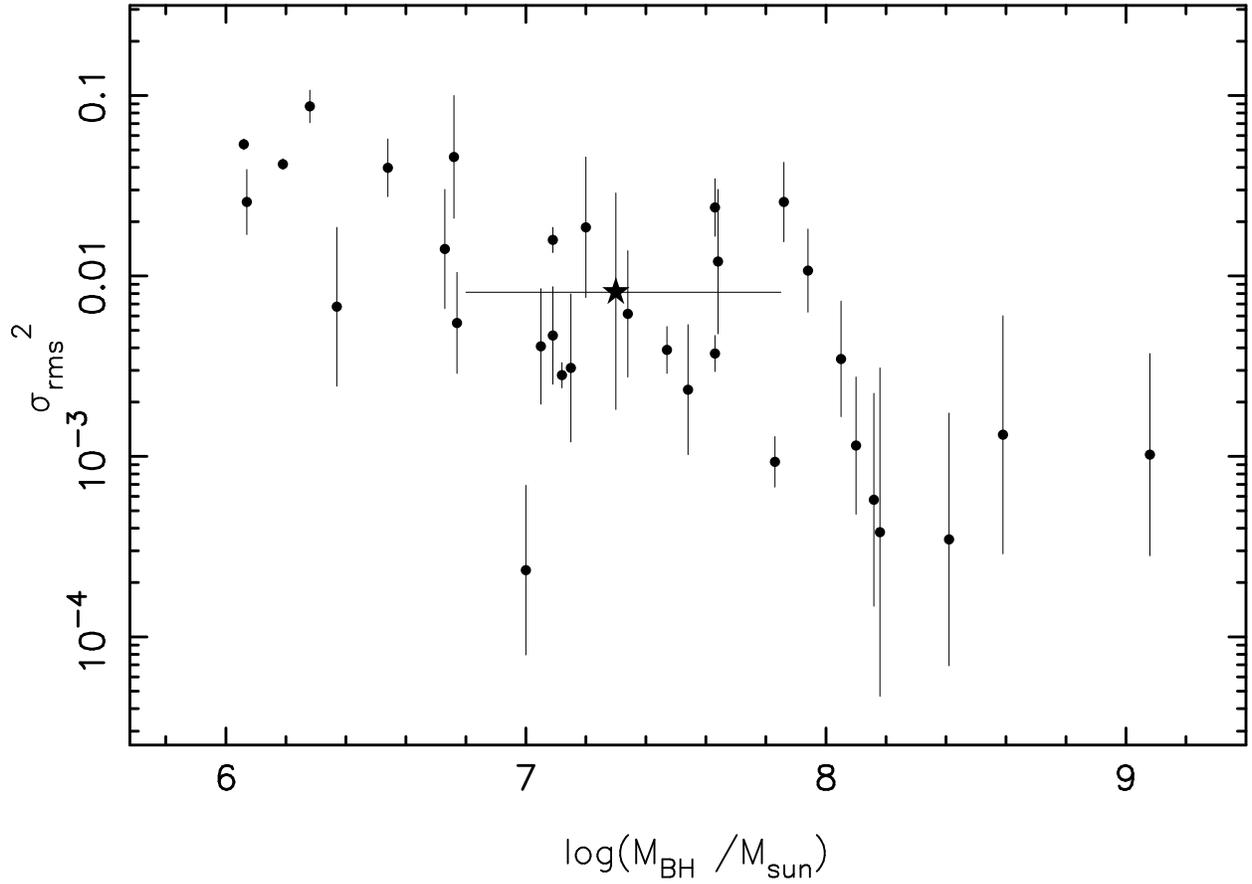}}
\figcaption{The X-ray excess variance $\sigma^2_{\rm rms}$ against the BH mass.
 The filled circles are taken from
O'Neill et al. (2005), the pentacle indicates our result of \he.}
\label{fig7}
\end{figure}
%

%
\begin{figure}
\figurenum{8}
\centerline{\includegraphics[angle=-90,width=\textwidth]{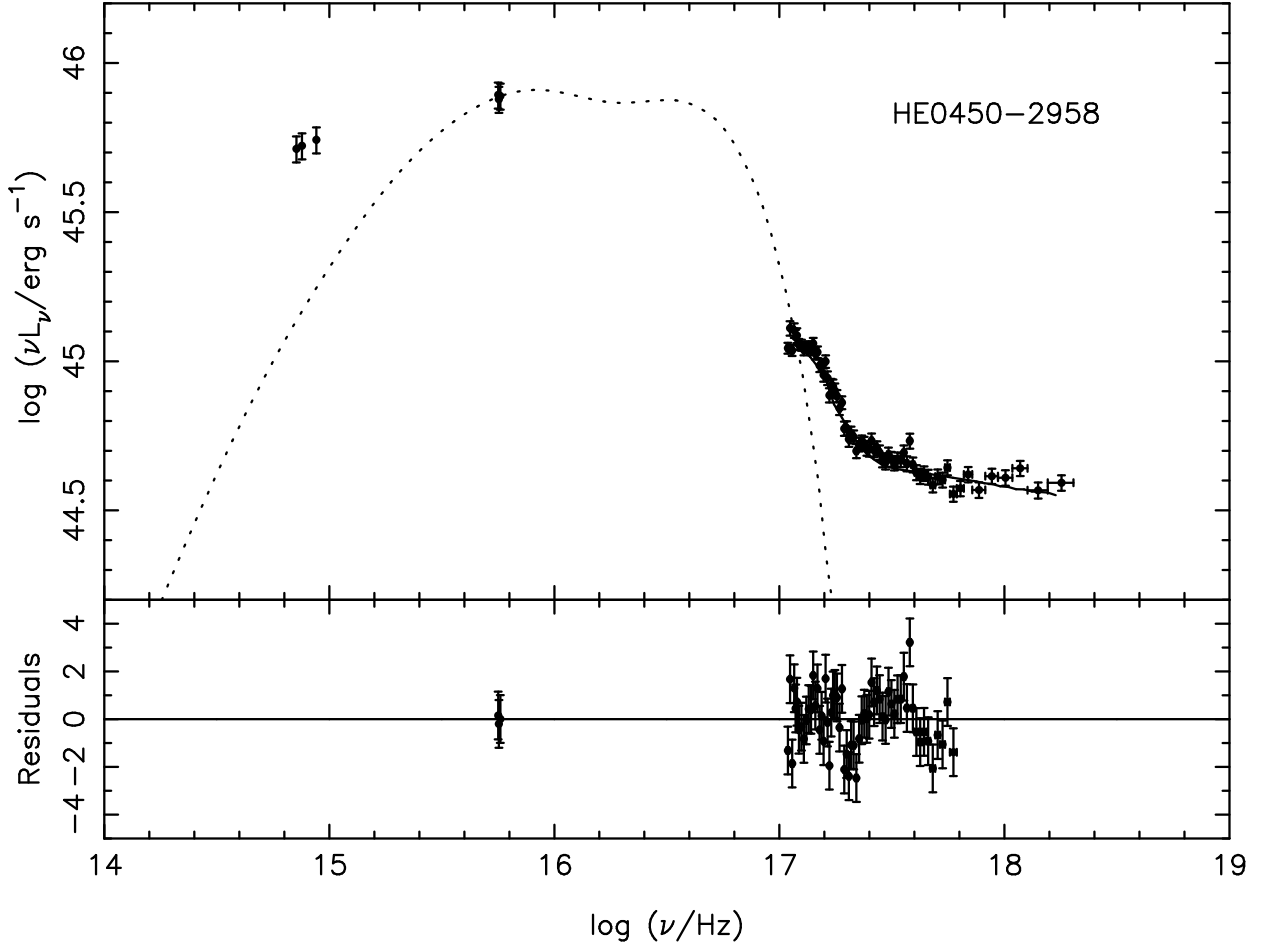}}
\figcaption{The spectral energy distribution of \he~ shows a big
blue bump in the UV band. The UV and optical data are taken from
Scott et al. (2004) and Letawe et al. (2006), respectively.  A
standard thin accretion disk model is used to fit the UV bump (dotted line). This gives
 a BH mass of $8\times10^{8}$ $M_{\odot}$ with an Eddington ratio of 0.3,
the inclination angle cos$\theta=0.5$ and the BH spin parameter
 $a=0.6$. This mass is much higher than the estimation
from the X-ray variability. The X-ray data are fitted with a power law plus a
blackbody model (solid line). }
\label{fig8}
\end{figure}
%


\begin{thebibliography}{}
\bibitem[]{659} Abramowicz, M. A., Czerny, B., Lasota, J. P., \& Szuszkiewicz,
E. 1988, \apj, 332, 646
\bibitem[]{661} Alexander, D. M., Smail, I., Bauer, F. E., Chapman, S. C.,
 Blain, A. W., Brandt, W. N., \& Ivison, R. J. 2005, Nature, 434, 738
\bibitem[]{663} Arnaud, K. A. 1996, ASPC, 101, 17
\bibitem[]{664} Ballantyne, D. R., Ross, R. R., \& Fabian, A. C.  2001, MNRAS, 323, 506
\bibitem[]{665} Barger, A. J., Cowie, L. L., Mushotzky, R. F., Yang, Y., Wang, W. H., Steffen, A. T., \& Capak, P.  2005, AJ, 129, 578
\bibitem[]{666} Bian, W.-H., \& Zhao, Y.-H. 2003, MNRAS, 343, 164
\bibitem[]{668} Boyce, P. J. et al. 1996, \apj, 473, 760
\bibitem[]{669} Braito, V. et al. 2004, \aap, 420, 79
\bibitem[]{670} Brandt, W. N., Mathur, S., \& Elvis, M.  1997, MNRAS, 285, L25
\bibitem[]{671} Canalizo, G., \& Stockton, A. 2001, \apj, 555, 719
\bibitem[]{672} Chapman, S. C., Blain, A. W., Ivison, R. J., \& Smail, I. R. 2003, Nature, 422,
695
\bibitem[]{674} Ciotti, L., \& Ostriker, J. P. 2001, \apj, 551, 131
\bibitem[]{C05} Crummy, J., Fabian, A. C., Gallo, L., \& Ross, R. R. 2006, MNRAS, 365, 1067
\bibitem[]{676} Dickey, J. M., \& Lockman, F. J. 1990, AR\aap, 28, 215
\bibitem[]{677} Di Matteo, T., Springel, V., \& Hernquist, L.  2005, Nature, 433, 604
\bibitem[]{678} D\"oerrer, T., Riffert, H., Staubert, R., \& Ruder, H.  1996, \aap, 311, 69
\bibitem[]{679} Elvis, M. et al. 1994, \apjs, 95, 1
\bibitem[]{680} Fabian, A. C.  1999, MNRAS, 308, L39
\bibitem[]{682} Gabriel, C. et al. 2004, ASPC, 314, 759
\bibitem[]{683} Gammie, C. F. 1998, MNRAS, 297, 929
\bibitem[]{684} George, I. M., Turner, T. J., Yaqoob, T., Netzer, H., Laor, A., Mushotzky, R. F.,
 Nandra, K., \& Takahashi, T.  2000, \apj, 531, 52
\bibitem[]{685} Ghizzardi S. et al. 2001, In flight calibration of the PSF for the MOS1 and
 MOS2 cameras, EPIC-MCT-TN-011
\bibitem[]{686} Gierli\'nski, M., \& Done, C. 2004, MNRAS, 349, L7
\bibitem[]{687} Haardt, F., \& Maraschi, L.  1993, \apj, 413, 507
\bibitem[]{688} Haehnelt, M. G., Davies, M. B., \& Rees, M. J.  2006, MNRAS, 366, L22
\bibitem[]{689} Hao, C.-N., Xia, X.-Y., Mao, S.-D., Wu, H., \& Deng, Z.-G.
2005, \apj, 625, 78
\bibitem[]{690} Imanishi, M., \& Terashima, Y. 2004, AJ, 127, 758
\bibitem[]{691} Hoffman, L., \& Loeb, A.   2006, \apj, 638, L75
\bibitem[]{692} Hopkins, P. F., Hernquist, L., Martini, P., Cox, T. J., Robertson, B.,
 Di Matteo, T., \& Springel, V. 2005, \apj, 625, L71
\bibitem[]{694} Kauffmann, G., \& Haehnelt, M. 2000, MNRAS, 311, 576
\bibitem[]{695} Kawaguchi, T., Pierens, A., \& Hur\'e, J. M. 2004, \aap, 415, 47
\bibitem[]{696} Kawakatu, N., Anabuki, N., Nagao, T., Umemura, M., \& Nakagawa, T.
2006, \apj, 637, 104
\bibitem[]{698} Letawe, G., Magain, P., Courbin, F., Jablonka, P., Jahnke, K., Meylan, G.,
\& Wisotzki, L. 2006, MNRAS(astro-ph/0605288)
\bibitem[]{700} Leighly, K. M. 1999, ApJS, 125, 297
\bibitem[]{701} Lu, Y. J., \& Yu, Q. J. 2001, MNRAS, 324, 653
\bibitem[]{702} Magain, P., Letawe, G., Courbin, F., Jablonka, P., Jahnke, K., Meylan, G.,
\& Wisotzki, L. 2005, Nature, 437, 381
\bibitem[]{MZ95} Magdziarz, P., \& Zdziarski, A. A. 1995, MNRAS, 273, 837
\bibitem[]{705} Malizia, A., Bassani, L., Stephen, J. B., Di Cocco, G., Fiore, F.,
\& Dean, A. J. 2003, \apj, 589, L17
\bibitem[]{707} Marconi, A., Risaliti, G., Gilli, R., Hunt, L. K., Maiolino, R., \& Salvati, M.
2004, MNRAS, 351, 169
\bibitem[]{709} Matt, G., Fabian, A. C., \& Ross, R. R. 1993, MNRAS, 262, 179
\bibitem[]{710} McLure, R. J., \& Dunlop, J. S. 2002, MNRAS, 331, 795
\bibitem[]{711} Merritt, D., Storchi-Bergmann, T., Robinson, A., Batcheldor, D., Axon, D., \&
Roberto, C. F. 2006, MNRAS, 367, 1746
\bibitem[]{713} Mineshige, S., Kawaguchi, T., Takeuchi, M., \& Hayashida,
K. 2000, PASJ, 52, 499
\bibitem[]{715} Mushotzky, R. F., Done, C., \& Pounds, K. A. 1993, AR\aap, 31, 717
\bibitem[]{716} Nandra, K., \& Pounds, K. A. 1994, MNRAS, 268, 405
\bibitem[]{717} Nandra, K., George, I. M., Mushotzky, R. F., Turner, T. J., \& Yaqoob, T.
1997, \apj, 476, 70
\bibitem[]{719} Nandra, K. 2006, MNRAS, 368, L62
\bibitem[]{720} Nayakshin, S., Kazanas, D., \& Kallman, T. R.  2000, \apj, 537, 833
\bibitem[]{721} Nayakshin, S., \& Melia, F. 1997, \apj, 484, L103
\bibitem[]{722} O'Neill, P. M., Nandra, K., Papadakis, I. E., \& Turner, T. J. 2005, MNRAS,
358, 1405
\bibitem[]{723} Papadakis, I. E. 2004, MNRAS, 348, 207
\bibitem[]{724} Piconcelli, E., Jimenez-Bailón, E., Guainazzi, M., Schartel, N.,
 Rodríguez-Pascual, P. M.,\& Santos-Lleó, M.  2005, \aap, 432, 15
\bibitem[]{726} Pounds, K. A., Done, C., \& Osborne, J. P.  1995, MNRAS, 277, L5
\bibitem[]{727} Pounds, K. A., Reeves, J. N., Page, K. L., Wynn, G. A., \& O'Brien, P. T. 2003,
MNRAS, 342, 1147
\bibitem[]{728} Pounds, K. A., \& Page, K. L. 2006, MNRAS, preprint (astro-ph/0607099)
\bibitem[]{729} Ross, R. R., \& Fabian, A. C. 2005, MNRAS, 358, 211
\bibitem[]{730} Reeves, J. N., \& Turner, M. J. L. 2000, MNRAS, 316, 234
\bibitem[]{731} Reynolds, C. S. 1997, MNRAS, 286, 513
\bibitem[]{SM96} Sanders, D. B., \& Mirabel, I. F. 1996, AR\aap, 34, 749
\bibitem[]{733} Scott, J. E., Kriss, G. A., Brotherton, M., Green, R. F., Hutchings, J., Shull, J.
 M., \& Zheng, W. 2004, \apj, 615, 135
\bibitem[]{735} Siemiginowska, A., Aldcroft, T. L., Bechtold, J., Brunetti, G., Elvis, M., \&
 Stanghellini, C. 2003, PASA, 20, 113
\bibitem[]{736} Shemmer, O., Brandt, W. N., Netzer, H., Maiolino, R., \& Kaspi, S. 2006,
\apj, 646, L29
\bibitem[]{737} Silk, J., \& Rees, M. J. 1998, \aap, 331, L1
\bibitem[]{738} Str\"uder, L. et al.  2001, \aap, 365, L18
\bibitem[]{739} Turner, M. J. L. et al.  2001, \aap, 365, L27
\bibitem[]{740} Turner, T. J., George, I. M., Nandra, K., \& Turcan, D. 1999, \apj, 524, 667
\bibitem[]{741} Turner, T. J., et al. 2002, \apj, 568, 120
\bibitem[]{744} Wang, J.-M., \& Netzer, H. 2003, \aap, 398, 927
\bibitem[]{745} Wang, J.-M., Szuszkiewicz, E., Lu, F.-J., \& Zhou, Y.-Y. 1999, \apj, 522, 839
\bibitem[]{746} Wang, J.-M., Watarai, K.-Y.,\& Mineshige, S. 2004, \apj, 607, L107
\bibitem[]{747} Zhou, X.-L, \& Wang, J.-M. 2005, \apj, 618, L83


\end{thebibliography}
\end{document}